\documentclass[fleqn,usenatbib]{mnras}

\usepackage{newtxtext,newtxmath}
\usepackage[T1]{fontenc}

\DeclareRobustCommand{\VAN}[3]{#2}
\let\VANthebibliography\thebibliography
\def\thebibliography{\DeclareRobustCommand{\VAN}[3]{##3}\VANthebibliography}

\usepackage{graphicx}	% Including figure files
\usepackage{amsmath}	% Advanced maths commands
\usepackage{soul,xcolor}
\setstcolor{red}

\title[Galactic distribution of pulsar scattering]{The Galactic distribution of pulsar scattering and the $\tau-{\rm DM}$ relation}

\author[Q.Y. He and X. Shi]{Qiuyi He,$^{1}$\thanks{E-mail: qiuyi981022@gmail.com}
Xun Shi$^{1}$\thanks{E-mail: xun@ynu.edu.cn}
\\
$^{1}$South-Western Institute for Astronomy Research (SWIFAR), Yunnan University, 650500 Kunming, P. R. China
}

\date{Accepted 2023 November 15. Received 2023 November 15; in original form 2023 August 30}

\pubyear{2024}

\begin{document}
\label{firstpage}
\pagerange{\pageref{firstpage}--\pageref{lastpage}}
\maketitle

\begin{abstract}
Interstellar radio wave scattering leads to flux density fluctuations and pulse broadening of pulsar signals. However, Galactic distribution and the structure of the scattering medium are still poorly understood. Pulsar pulse broadening data available for a relatively large number of pulsars is well suited for such investigations. We collected an up-to-date sample of publicly available pulsar scattering data and introduced a new quantity -- the reduced scattering strength $\tilde{\tau}$ to study the Galactic distribution of pulsar scattering in the Milky Way. We show that the current observations are dominated by two distinct pulsar populations: a local and an inner-Galactic one separated by $\tilde{\tau }=10^{-5.1}$ \, s\, cm$^{6}$\,pc$^{-1}$. The stronger electron density fluctuations associated with the inner-Galactic population naturally explain the observed steepening of pulsar scattering time $\tau$ - dispersion measure relation. We measure an inner disc region with $3\,{\rm kpc}<\rm r< 5.5\,{\rm kpc}$ from the Galactic centre to have a scattering scale height of about 0.28\,kpc, supporting a correlation between interstellar radio scattering and  structures associating with the ionized gas and stellar activities. 
\end{abstract}


%%%%%%%%%%%%%%%%%%%%%%%%%%%%%%%%%%%%%%%%%%%%%%%%%%

%%%%%%%%%%%%%%%%% BODY OF PAPER %%%%%%%%%%%%%%%%%%

\section{Introduction}
\label{Se:In}

After the initial discovery of pulsars, it was soon realized that their flux density exhibits fluctuations from minutes to months \citep{Backer75, cordes76}. These variations were subsequently attributed to the scattering effect caused by unevenly distributed interstellar plasma interfering with the radio waves passing through \citep{Bonazzola75, Isaacman77}. In addition to the flux density changes, this phenomenon, known as pulsar scintillation, is also responsible for the widening of pulse profiles. The scattering occurring within the interstellar medium introduces a time delay to the redirected radio waves, resulting in the observed broadening effect \citep{Armstrong81, Cordes91}.

The effect of scattering-induced pulse broadening of the original pulse profile is described by the convolution with a pulse broadening function that is usually assumed to have a simple exponential form $\mathrm{PBF(\mathit{t})\sim exp(-\mathit{t/\tau})}$ \citep{7.1968Natur.218..920S}, where $\tau$ is the pulse spreading time (also called the scattering time). It has been found that the scattering time $\tau$ is smaller at higher observing frequency \citep{15.2004A&A...425..569L,16.2013MNRAS.434...69L}. This is because, at higher frequencies, the phase perturbation in the radio wave is smaller when it passes through the interstellar medium, leading to a smaller deflection angle and thus smaller time delay. 

Scintillation studies that utilize scattering time and flux variations are highly complementary to each other \citep{55.2011AIPC.1357...97R}. While scattering time quantifies the overall scintillation intensity, flux variations as a function of time and frequency i.e. the dynamic spectra offer insights into the time-scale and frequency distribution, providing detailed information about the scattering medium. However, dynamic spectrum observations are considerably more time consuming and demand high telescope sensitivity or strong pulsar signals. Scattering data is more appropriate and suitable for statistical investigations of interstellar scattering within the Milky Way.

Based on scattering data, the pulse broadening time $\tau$ caused by pulsar scintillation is found to correlate positively with the dispersion measure (DM; e.g. \citealt{Sutton71, 29.1997MNRAS.290..260R, 17.2004ApJ...605..759B, 25.2015ApJ...804...23K, 6.2016arXiv160505890C}), Notably, this correlation exhibits different power-law slopes for low-DM and high-DM pulsars, with a transition occurring around DM $\approx 40$\, pc\, cm$^{-3}$. While the steepening of the slope at higher DM values is generally expected to be associated with increased scattering towards the inner regions of the Galaxy, a clear and comprehensive understanding is still lacking.

In this paper, we explore the Galactic distribution of pulsar scattering that underlies the apparent $\tau-$DM relation in terms of a newly defined `reduced scattering strength' $\tilde{\tau}$,  using an up-to-date set of scattering data collected from the literature. After explaining our data collection in Section~\ref{sc:da}, we introduce $\tilde{\tau}$ in Section~\ref{sc:reduce} and show that its Galactic distribution reveals two distinct pulsar populations. In Section~\ref{sc:relation}, we re-interpret the $\tau$-DM relationship using the two pulsar populations. In Section~\ref{sc:hi} we constrain the vertical scale height of scattering in the inner-disc region. Finally, we conclude in Section~\ref{sc:con}.

\section{Data collection}
\label{sc:da}

Our data collection is based on the publicly available pulsar scattering time data from the ATNF website (Catalogue Version: 1.70). The scattering time values provided on the website are all normalized to 1GHz observation frequency assuming $\tau \propto \nu^{-4.4}$. Given the wide range of observing frequency of pulse broadening measurements (from below 40 MHz to above 2000 MHz) and the dispersion of the actual frequency scaling of $\tau$ \citep{15.2004A&A...425..569L, 36.2015MNRAS.454.2517L, 5.2019ApJ...870....8S}, normalizing using this steep relation could potentially introduce large errors in the data points. Therefore, we retain $\tau$ values at all observed frequencies in our data collection but also provide $\tau$ values normalized to 1 GHz using $\tau \propto \nu^{-4}$ following \citet{6.2016arXiv160505890C}. In this paper, we use this frequency-normalized value as a default and discuss the validity of the frequency-normalization in Appendix~\ref{A}. 

We traced back to the original references and obtained the corresponding scattering time values at the observed frequencies. Measurements at multiple frequencies were all included when available. We also supplemented our data collection by searching for additional pulsar observation data mentioned in each reference. Next, we included additional data mentioned in \citet{6.2016arXiv160505890C}. Finally, we added results from a recently published article \citep{10.2022arXiv220708756S} and some articles not included on the ATNF website \citep{69.2017MNRAS.470.2659G, 70.2017ApJ...846..104K, 76.2019ApJ...878..130K}. 

We finally formed a data base with 1110 pulsar scattering time observations obtained by measuring 473 pulsars at different observation frequencies from 54 papers. This is the most complete multifrequency data of $\tau$ measurements compiled to date. Among them, 244 pulsars have  scattering time data at a single frequency, and 229 pulsars have multifrequency data (see Appendix~\ref{B}). The DM and distance values for the sample were taken from the ATNF data base. For pulsars with parallax distances, we replaced this default distance derived from the pulsar DM using the YMW16 electron density model with the parallax distance, and mark these pulsars with `yes' in the `parallax' column.

\section{The reduced scattering intensity $\tilde{\tau}$}
\label{sc:reduce}

\begin{figure}
	\includegraphics[width=8cm]{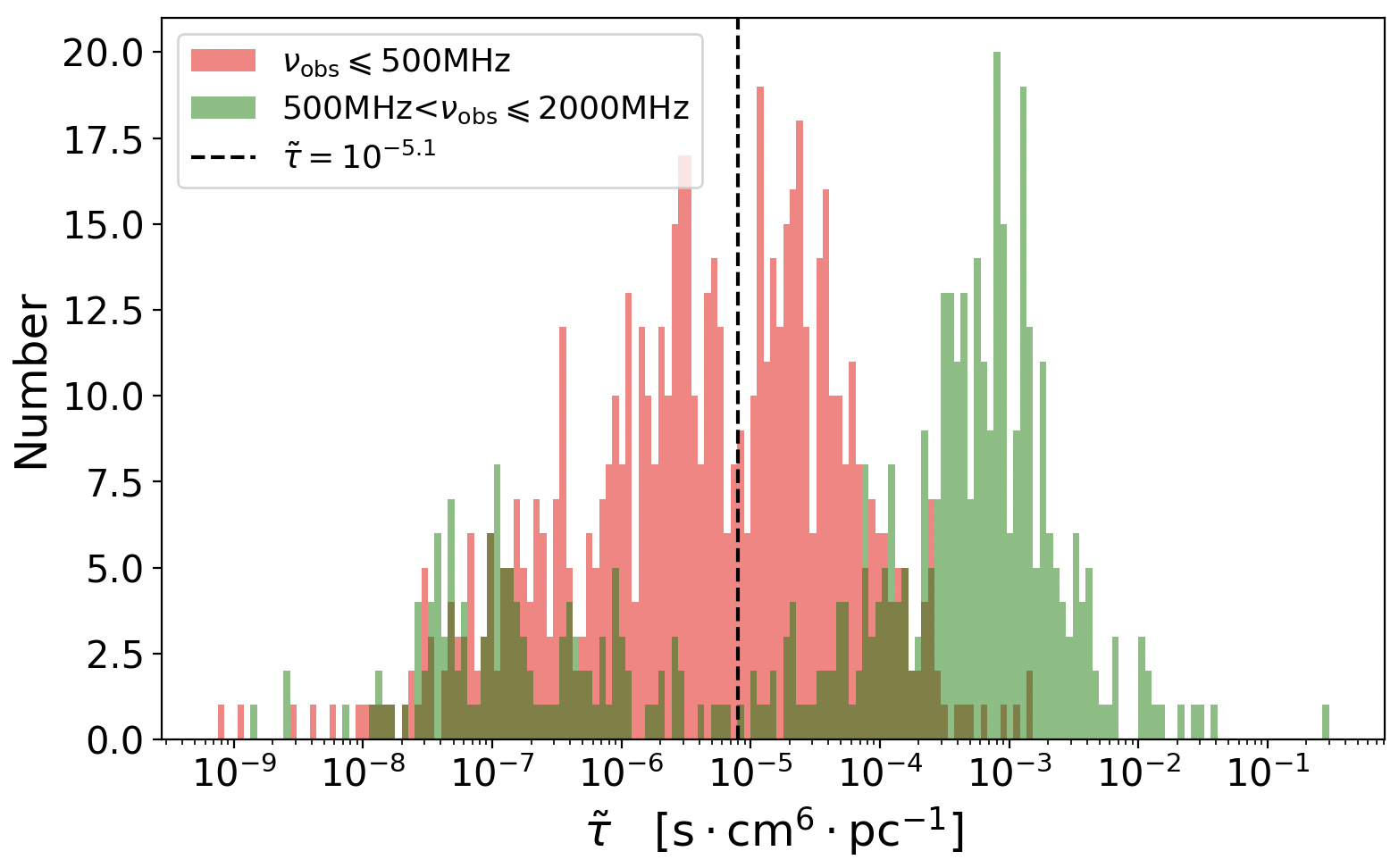}
    \caption{Distribution of the sample over the reduced scattering intensity $\tilde{\tau }$. For both low and high-frequency observations, the number of measurements around $\tilde{\tau }=10^{-5.1}$ \, s\, cm$^{6}$\,pc$^{-1}$ is relatively small, leading to a bimodal distribution of $\tilde{\tau}$.} 
    \label{Data:Distribution}
\end{figure}

\begin{figure}
	\includegraphics[width=8.3cm]{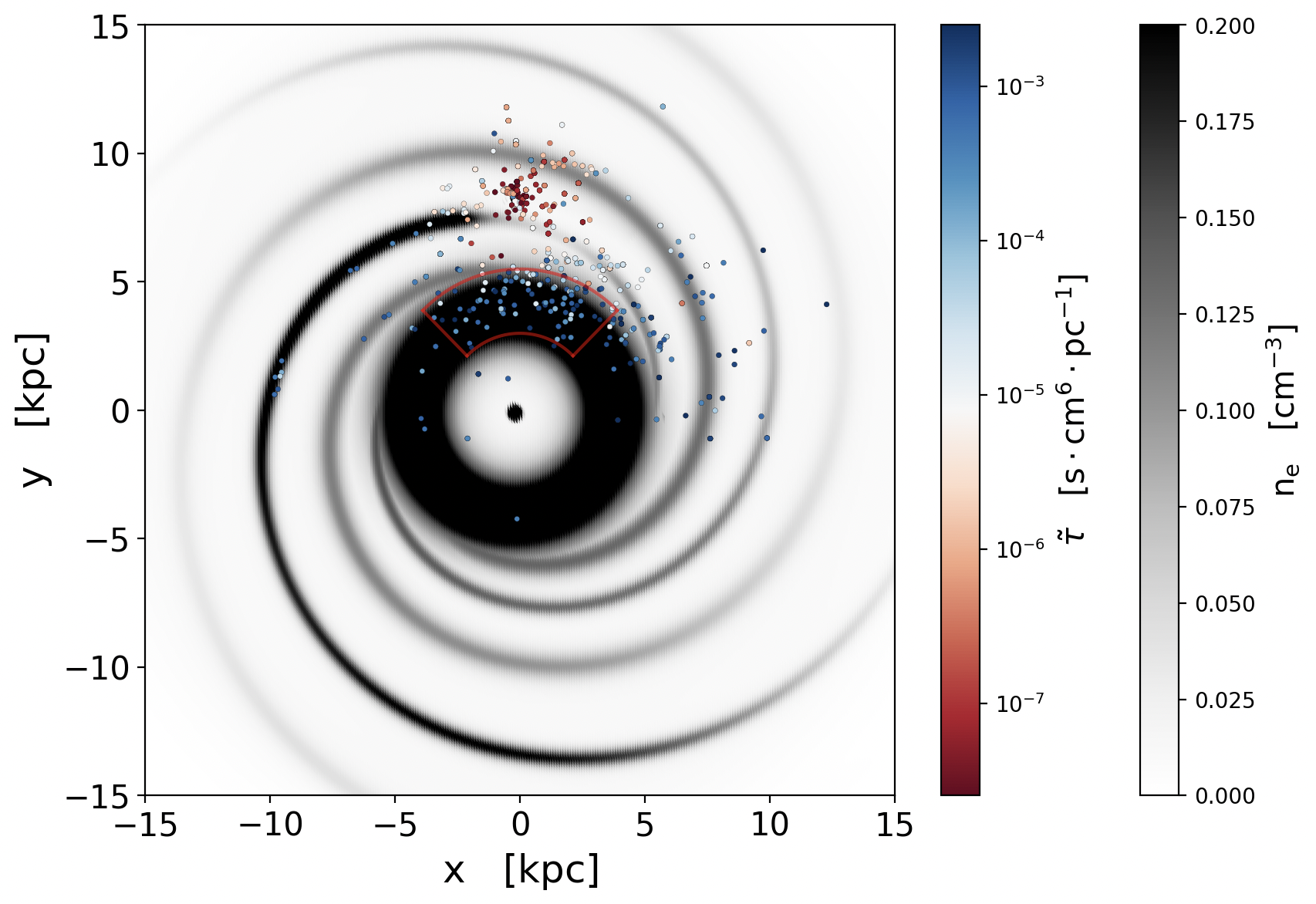}
    \caption{The Galactic distribution of low Galactic latitude pulsars (the vertical distance to the plane of the galaxy is less than 0.4\,kpc) color-coded by their reduced scattering strength $\tilde{\tau}$ superimposed on the mid-plane Galactic electron density distribution given by the YMW16 model. The two peaks in the value of $\tilde{\tau}$ (Fig.\;\ref{Data:Distribution}) are clearly dominated by two pulsar populations. The data points in the red fan-like region will be used for the scattering scale-height analysis in Section~\ref{sc:hi}.}
    \label{sc:allymw}
\end{figure}

Physically, the DM represents the integrated free electron density  $n_{\rm e}$ in the pulsar line of sight, $\mathrm{DM}\propto \int_{0}^{D} n_{\rm e}\rm{d} \ell = \mathit{\left \langle n_{\rm e}\right \rangle D}$, where $D$ is the distance to the pulsar. Accordingly, the scattering time is proportional to the integrated electron density fluctuation, $\tau \propto \int_{0}^{D} \left (n_{\rm e}^2 -\bar{ n_{e}} ^{2} \right ) \rm{d} \ell = \mathit{ \left\langle {\mathrm{\Delta}} n_{e}^2 \right\rangle D}$. In both quantities, there is a degeneracy between distance and electron density/density fluctuation. We will show that this degeneracy explains part of the large scatter in the $\tau-$DM relation. In an effort to overcome this degeneracy and to focus on the scattering, we introduce a combined quantity $\tilde{\tau }=\frac{\tau \cdot D}{{\rm DM}^{2}}\sim \frac{\left \langle {\mathrm{\Delta}} {n_{\rm e}}^{2}\right \rangle}{\left \langle n_{\rm e}\right \rangle^{2}}$, which we refer to as the reduced scattering intensity. This reduced scattering directly reflects the reduced electron density fluctuation and can be related to the fluctuation parameter in Galactic scattering models \citep{54.2002astro.ph..7156C, 9.2003astro.ph..1598C}.

Fig.~\ref{Data:Distribution} shows the distribution of $\tilde{\tau}$ for both low-frequency ($\nu_{\rm obs} \le 500$ MHz) and high-frequency (500 MHz $< \nu_{\rm obs} \le 2000$ MHz) observations in our sample. Regardless of the observing frequency, the number of measurements around $\tilde{\tau}=10^{-5.1}$ \,s\,cm$^{6}$\,pc$^{-1}$ is relatively small, separating two distinct populations at lower and higher $\tilde{\tau}$ values.

The Galactic distribution of $\tilde{\tau}$ confirms the existence of two main pulsar populations underlying the current scattering data. Fig.~\ref{sc:allymw} shows the distribution of the `Galactic-plane' pulsars (those with distances to the Galactic plane $|\rm{Z}|<0.4$\,kpc, which constitute 77.2 per cent of our whole sample) superimposed on the mid-plane Galactic electron number density distribution given by the YMW16 model \citep{18.2017ApJ...835...29Y}. We can clearly see that the pulsars with different reduced scattering intensities are clustering in different locations in the Galaxy. While pulsars with $\tilde{\tau} < 10^{-6.1}$ \, s\, cm$^{6}$\,pc$^{-1}$ (red points) form a local population in the vicinity of Earth, pulsars with $\tilde{\tau } > 10^{-4.1}$ \, s\, cm$^{6}$\,pc$^{-1}$ (blue points) reside on spiral arms and pre-dominantly, a region in the inner Galaxy referred to as the `inner thin disc' (thick black ring in Fig.~\ref{sc:allymw}) in the electron density models \citep{54.2002astro.ph..7156C, 9.2003astro.ph..1598C, 18.2017ApJ...835...29Y}.

Note that, according to our definition of the reduced scattering intensity $\tilde{\tau}$, there is no apparent distance dependence in it. Thus, the overall trend that pulsars with higher $\tilde{\tau}$ reside further away from us is \textit{not} trivial. It implies that the fractional electron density fluctuations in our local environment are much smaller than those in the inner Galaxy. If an observer were to study the identical group of pulsars from a position within the inner disc, they would find that the more distant pulsars in Earth's vicinity would still exhibit smaller values for $\tilde{\tau}$ compared to the `local' pulsar population situated within the inner disc. This is the case despite the fact that the more distant pulsars would show larger ${\tau}$ values.

\section{$\tau-{\rm DM}$ relation}
\label{sc:relation}

\begin{figure}
	\includegraphics[width=8.5cm]{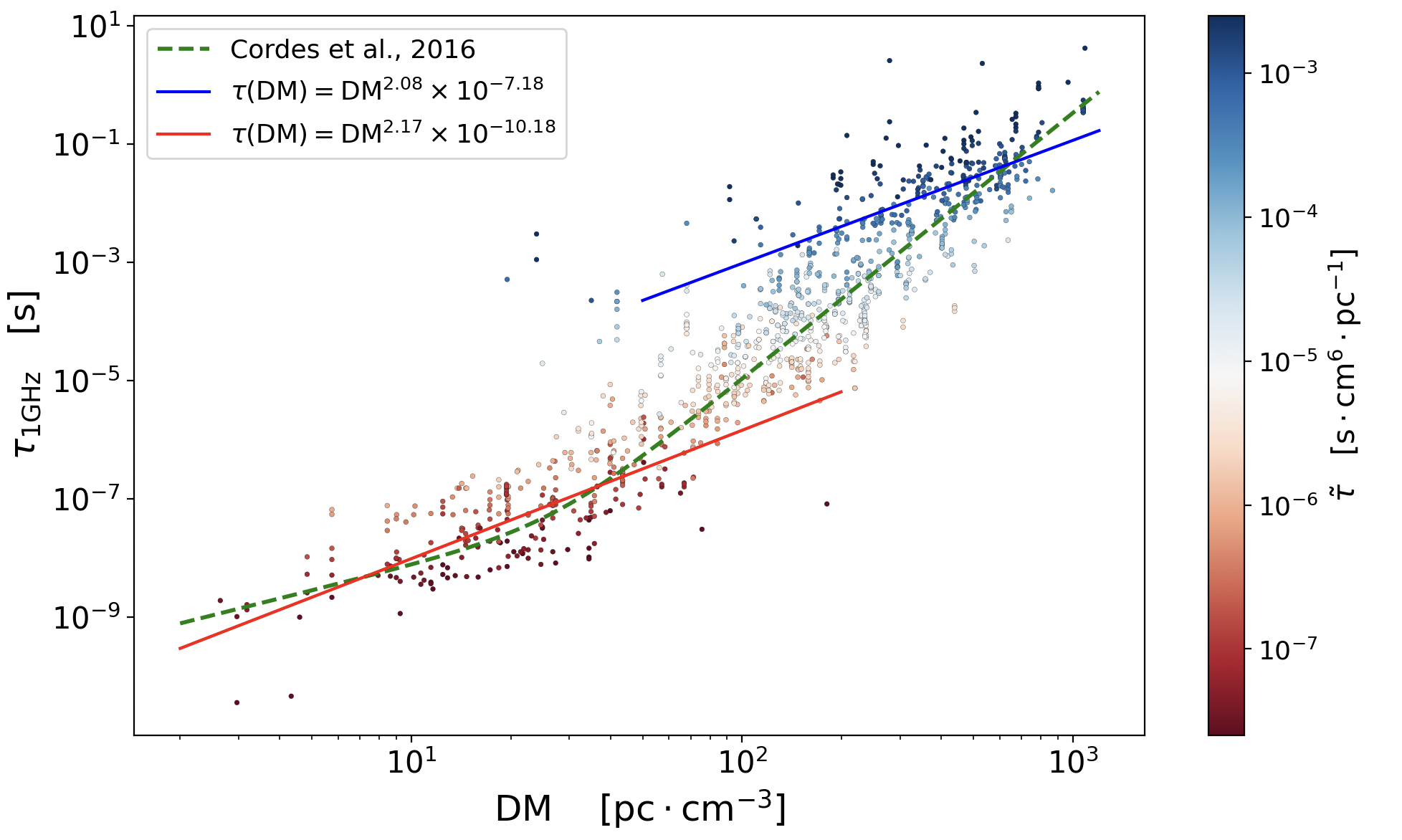}
    \caption{Scatter plot of the scattering time delay $\tau$ versus dispersion measure DM using our sample. The fitting curve of \citealt{6.2016arXiv160505890C} (green dashed line) is consistent with the enlarged sample. Nevertheless, we suggest fitting $\tau-$DM relation for individual pulsar samples. The two pulsar populations with $\tilde{\tau } < 10^{-6.1}$ \, s\, cm$^{6}$\,pc$^{-1}$ and $\tilde{\tau } > 10^{-4.1}$ \, s\, cm$^{6}$\,pc$^{-1}$ have power-law $\tau-$DM relations with very different amplitudes (red and blue line, respectively).} 
    \label{Data:ourresult}
\end{figure}

Fig.~\ref{Data:ourresult} shows the relationship between the scattering time $\tau$ and the $\rm DM$ plotted using our sample. Despite the nearly doubled number of measurements, the data points match well with the \citet{6.2016arXiv160505890C} fitting curve (green dashed line), showing the consistency of the data sets.

\begin{figure}
	\includegraphics[width=8.5cm]{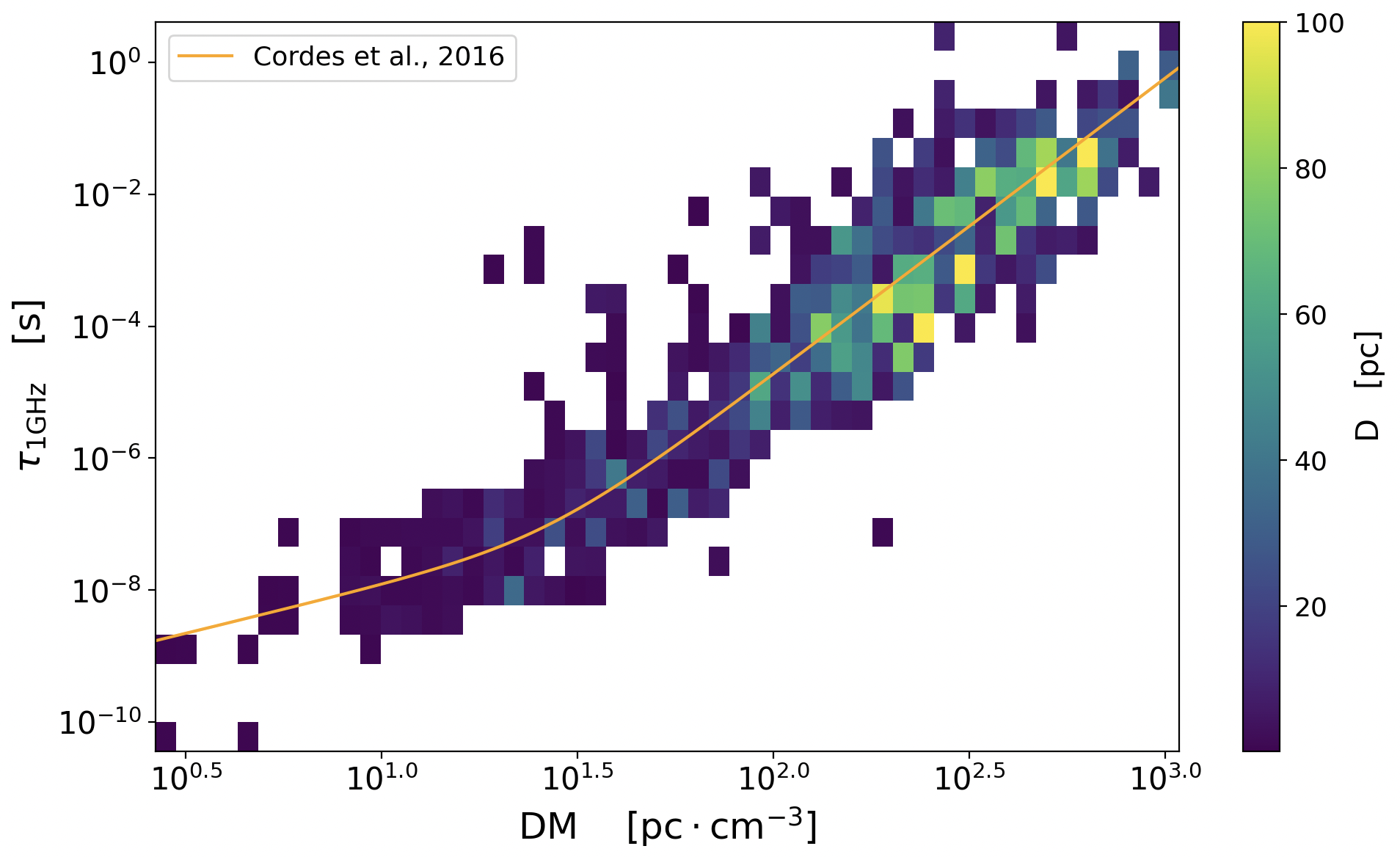}
    \caption{Meaurements lying below the average $\tau-$DM relation (the orange line) are from pulsars at larger average distances. This demonstrates the degeneracy between distance and electron density in the $\tau-$DM relation which partly explains the large scatter in the $\tau-$DM relation and motivates the use of the reduced scattering strength $\tilde{\tau}$.}
    \label{fi:bin}
\end{figure}

The location segregation of the reduced scattering intensity (Fig.~\ref{sc:allymw}) implies substantially different electron density distributions in different regions in our Galaxy. This motivates us to fit the $\tau-$DM relation separately for the two distinct pulsar populations. 

We select samples with $\tilde{\tau}$ values greater than $10^{-4.1}$ and less than $10^{-6.1}$ as the local and inner-Galaxy populations, respectively, and fit a power-law relation $\tau (\rm DM)={\rm DM}^{\it a}\times 10^{\it b}$ to each population. We used the \texttt{LMFIT} package for fitting \footnote{https://lmfit.github.io/lmfit-py/}, and in the Appendix~\ref{C}, we provided the parameter covariance obtained from the \texttt{EMCEE} sampling process, visualized using the \texttt{CORNER} package \citep{lmfit}. We similarly apply this method to the scattering scale height of the inner disc discussed in Section~\ref{sc:hi}. The resulting fitting parameters read
\begin{equation}
\begin{matrix}
a_{\rm inner}=2.08 \pm 0.10, b_{\rm inner}=-7.18 \pm 0.27,\\ 
a_{\rm local}=2.17 \pm 0.09, b_{\rm local}=-10.18 \pm 0.13.\nonumber
\end{matrix}
\end{equation}
It can be seen that although the two populations have very different amplitudes, they have similar slopes. For the inner-Galaxy population, the $\tau-$DM relation (blue line) shows a much higher amplitude than the local population, as expected. 

The large scatter of the $\tau-$DM relation is partly due to the distance-electron density/electron density fluctuation degeneracy in DM and $\tau$. This degeneracy implies, for example, that a far-away high-latitude pulsar can have the same DM as a nearby pulsar on the Galactic plane. However, the latter usually has a much longer scattering time than the former, leading to a large spread on the $\tau-{\rm DM}$ relation. Fig.~\ref{fi:bin} illustrates this by showing the average distance of the underlying pulsars on the $\tau-{\rm DM}$ plane. Pulsars lying below the mean $\tau-{\rm DM}$ relation indeed tend to be located at larger distances. This is in accordance with the finding that fast radio bursts (FRBs) are under-scattered compared with Galactic pulsars of comparable DM \citep{6.2016arXiv160505890C, Cordes22, Ravi19}.

\section{Scattering scale height of the inner disc} 
\label{sc:hi}

What is the interstellar structure responsible for the scattering of pulsars is still a mystery. Apart from the association with the Local Bubble structures \citep{Bhat16, Reardon20, McKee2022, Yao22, Liu23}, a number of ideas have been proposed, including \ion{H}{II} regions \citep{64.2000ApJ...539..300S, Mall22}, superbubble shells \citep{10.2022arXiv220708756S}, supernova remnants \citep{65.2020ApJ...897..124O, 2021Yao}, ionized skins of molecular clumps \citep{walker17}, local interstellar clouds \citep{McKee2022}, filaments ionized by hot stars \citep{walker17}, and corrugated plasma reconnection sheets \citep{pen12, pen14b, liu16, simard18}. These proposed astrophysical sources have different vertical distributions over the Galactic plane. Thus,  one effective way to constrain the underlying astrophysical source of scattering is to measure the vertical distribution of scattering. 

\begin{figure}
   \centering
   \includegraphics[width=8.3cm]{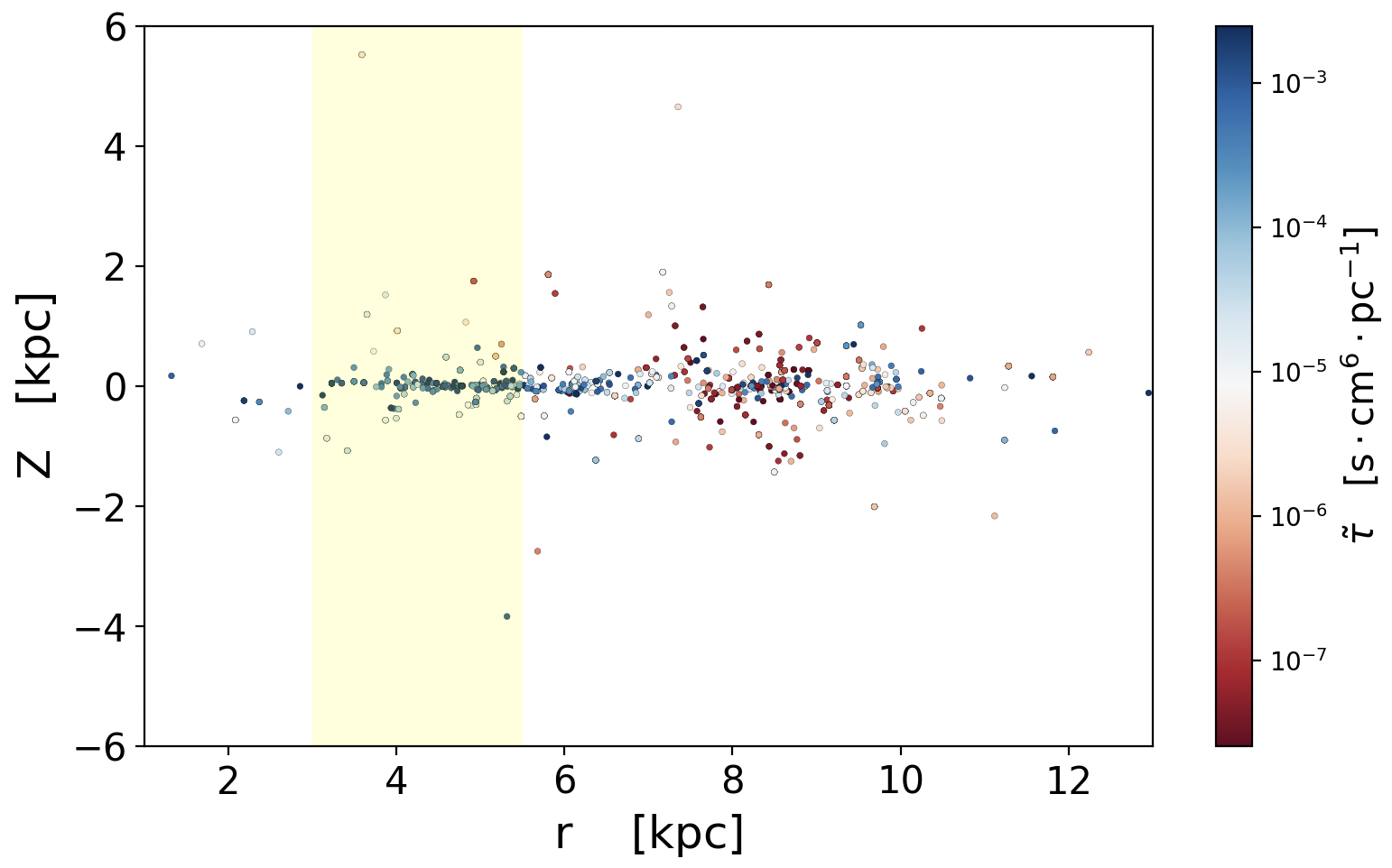}
   \caption{Galactic $r-$Z distribution of our pulsar sample. Pulsars with distances to the Galactic centre in the range of 3\,kpc $<r<$5.5\,kpc (yellow shaded region) have reduced scattering intensities (color-coding) that show little dependence on $r$, allowing us to select a homogeneous `inner disc' population from this region.}
   \label{hidata}
\end{figure}

\begin{figure*}
	\includegraphics[width=\linewidth]{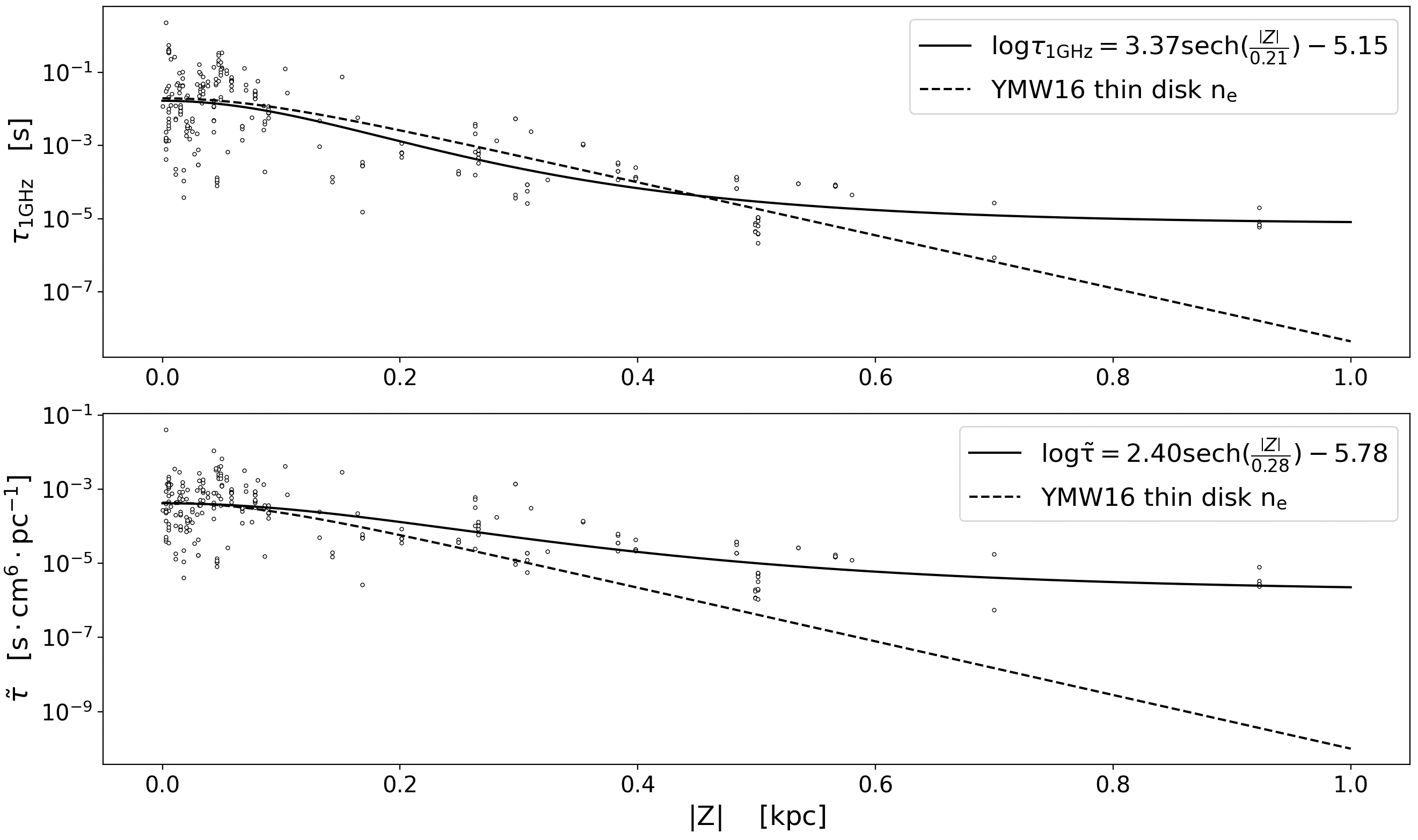}
    \caption{Pulse broadening time (upper panel) and reduced scattering strength (lower panel) distribution over distance to the Galactic plane for pulsars in the  selected inner disc region (the red fan-shaped region in Fig.\;\ref{sc:allymw}) where scattering is predominantly dependent on $|\rm{Z}|$. Our fits are given by the solid lines. As a comparison, the dashed lines show the $|\rm{Z}|$-dependence of the thin-disc electron number density in the YMW16 model (scaled to share the same value with the solid lines at $|\rm{Z}|=0$\;kpc.)}
    \label{hifit}
\end{figure*}

To obtain a homogeneous sample for such analysis, We plot the $r-\rm Z$ distribution of the reduced scattering strength of our pulsar sample (Fig.~\ref{hidata}), where $r = \sqrt{\rm{X}^2 + \rm{Y}^2}$ and X, Y, Z are the Cartesian Galactic coordinates centred on the Galactic centre. We find that the reduced scattering strength in the inner disc region at 3\,kpc $< r <$ 5.5\,kpc (yellow region) is relatively independent of $r$. Within this radial range, we select pulsars in a compact region between us and the Galactic centre (the red fan-like region on Fig.~\ref{sc:allymw} with Y$>$0 and $|\rm{Y/X}|>1$) as our sample for analysing the vertical scattering structure of the inner disc. We further limit the analysis to $\left | \rm Z\right |\leqslant 1 \, {\rm kpc}$ where the distributions above and below the Galactic plane are symmetric and well sampled.

Fig.~\ref{hifit} presents the vertical distribution of the corresponding scattering data in terms of pulse broadening time $\tau_{\rm 1GHz}$ and the reduced scattering intensity $\tilde{\tau}$. The data points at a certain distance $|\rm{Z}|$ have a large spread in their scattering strength, showing the strong line-of-sight dependence of scattering. This is consistent with the current picture that the scattering is caused by small-scale electron density inhomogeneities mainly contributed by a small number of objects. Nevertheless, there is a clear trend of a decreasing average scattering strength with $|\rm{Z}|$. We fit this using a form commonly used for vertical distributions on the Galactic disc:
\begin{equation}
    \mathrm{log \mathit{K} = \mathit{C} sech(\frac{\left |\rm{Z} \right |}{\mathit{h}})+\mathit{c}}.
	\label{eq:KZ}
\end{equation}
In this equation, $K$ represents either the reduced scattering intensity or the pulse broadening at 1\,GHz, and $h$ is the vertical scale height. The resulting fitting parameters are
\begin{equation}
\begin{matrix}
C_{\tau_{1 \mathrm{GHz}}} = 3.37 \pm 0.31, h_{\tau_{1 \mathrm{GHz}}} = 0.21 \pm 0.03, c_{\tau_{1 \mathrm{GHz}}} = -5.15 \pm 0.32. \nonumber \\
C_{\tilde{\tau}} = 2.40 \pm 0.34, h_{\tilde{\tau}} = 0.28 \pm 0.05, c_{\tilde{\tau}} = -5.78 \pm 0.35. \nonumber
\end{matrix}
\end{equation}

Scale heights of several gas and stellar discs are well-measured: the scale height of the \ion{H}{I} disc in the inner region of the Milky Way is approximately 0.1\,kpc \citep{66.2009ARA&A..47...27K}, that of the molecular gas in the Milky Way is approximately 50\,pc \citep{67.2022MNRAS.515.1663J}, and that of OB stars is also about 50\,pc. Although these neutral, relatively dense gases can easily provide sufficient electron density fluctuations for the scattering when they are ionized by e.g. star formation activities, their scale heights are far smaller than that of the scattering medium, which we measure to be around 0.28\,kpc.  On the other hand, the scale height of the `thin' stellar disc has been reported to be around 0.3\,kpc \citep{68.2013A&ARv..21...61R}. The electron density distribution in the thin disc components in the   Galactic electron number density distribution models NE2001 \citep{54.2002astro.ph..7156C, 9.2003astro.ph..1598C} and YMW16 \citep{18.2017ApJ...835...29Y} also shares a similar scale height (see Fig.\;~\ref{hifit}). These imply, if a single type of structure is responsible for the  distribution of interstellar radio wave scattering, it is possibly associated with the ionized gas and/or stellar activities, but not with neutral gas, molecular gas, or star formation.

\section{Conclusions}
\label{sc:con}

Studying the mechanism and characteristics of pulsar pulse broadening can help us better understand the distribution and nature of the radio wave scattering medium in the Galaxy. We investigate the relationship between $\tau$ and $\rm DM$ based on the three-dimensional structure of the ionized gas distribution in the Galaxy and design the reduced scattering intensity $\tilde{\tau}$ -- a combination of measurable quantities that is directly proportional to $\left \langle {\mathrm{\Delta}} n_{\rm e}^2\right \rangle / {\left \langle n_{\rm e}\right \rangle}^2$. As a measure of the scattering strength, the reduced scattering intensity has no apparent dependence on the pulsar distance or the average number of electrons in the path. Thus, it can help us analyze more objectively the effect of the structure of the various parts of the Galaxy on the scintillation of the pulsar.

We have assembled an up-to-date sample of publicly available pulsar scattering observation data, which includes multifrequency data for each pulsar. The sample reveals two pulsar populations according to the magnitude of their reduced scattering intensity $\tilde{\tau}$, separated by $\tilde{\tau }=10^{-5.1}$ \, s\, cm$^{6}$\,pc$^{-1}$. Their Galactic distribution confirms this bimodality, that the two populations correspond well to a local, and an inner-Galactic population, respectively.

Viewing the $\tau-$DM relation in the light of these two pulsar populations, we obtained power-law fitting relations for these two pulsar populations: $\tau ({\rm DM_{inner}}) = {\rm DM}^{2.08 \pm 0.10} \times 10^{-7.18 \pm 0.27}$ and $\tau ({\rm DM_{local}}) = {\rm DM}^{2.17 \pm 0.09} \times 10^{-10.18 \pm 0.13}$, suggesting that the radio waves from pulsars in the inner-Galactic population propagate through a medium with a much higher electron density fluctuation amplitude. When the high-DM values are not primarily dominated by the inner disc population e.g. in the case of high-latitude FRBs, the $\tau-{\rm DM}$ relation would exhibit different characteristics.

A significant fraction of the  inner-Galactic pulsar scattering data belongs to an inner-disc region within $3\,{\rm kpc}<\rm r< 5.5\,{\rm kpc}$ from the Galactic centre. Using this sample, we measure the vertical scale height of the pulse broadening and the reduced scattering intensity of the inner Galactic disc as 0.21\,kpc and 0.28\,kpc, respectively. This suggests that if there is a specific structure responsible for the scattering of interstellar radio waves, it may be connected to ionized gas and/or stellar activities, rather than neutral gas, molecular gas, or star formation. 

In the future, through further pulsar surveys and expanding the data set of scattering measurements, we can gain a more comprehensive understanding of the electron density fluctuations within the Milky Way. This will enable us to explore the complex structures of the interstellar medium and their impact on various astrophysical phenomena. Additionally, these studies will have practical applications in e.g. pulsar timing where accurate measurements of scattering properties will be very helpful.

%%%%%%%%%%%%%%%%%%%%%%%%%%%%%%%%%%%%%%%%%%%%%%%%%%

\section*{Acknowledgements}
We thank Wojciech Lewandowski for explaining the data in their article and for useful discussions. We thank the anonymous referee for the valuable feedback and suggestions.
This project is supported by NSFC No. 12373025.

\section*{Data Availability}
The data supporting the findings of this study are available within this paper and its supplementary information files.



\bibliographystyle{mnras}
\bibliography{references} % if your bibtex file is called example.bib

%\begin{thebibliography}{99}
%\end{thebibliography}

%%%%%%%%%%%%%%%%%%%%%%%%%%%%%%%%%%%%%%%%%%%%%%%%%%

%%%%%%%%%%%%%%%%% APPENDICES %%%%%%%%%%%%%%%%%%%%%

\appendix

\section{Effect of observing frequency}
\label{A}

\subsection{Frequency dependence of $\tau$}

\begin{figure}
	\includegraphics[width=8.3cm]{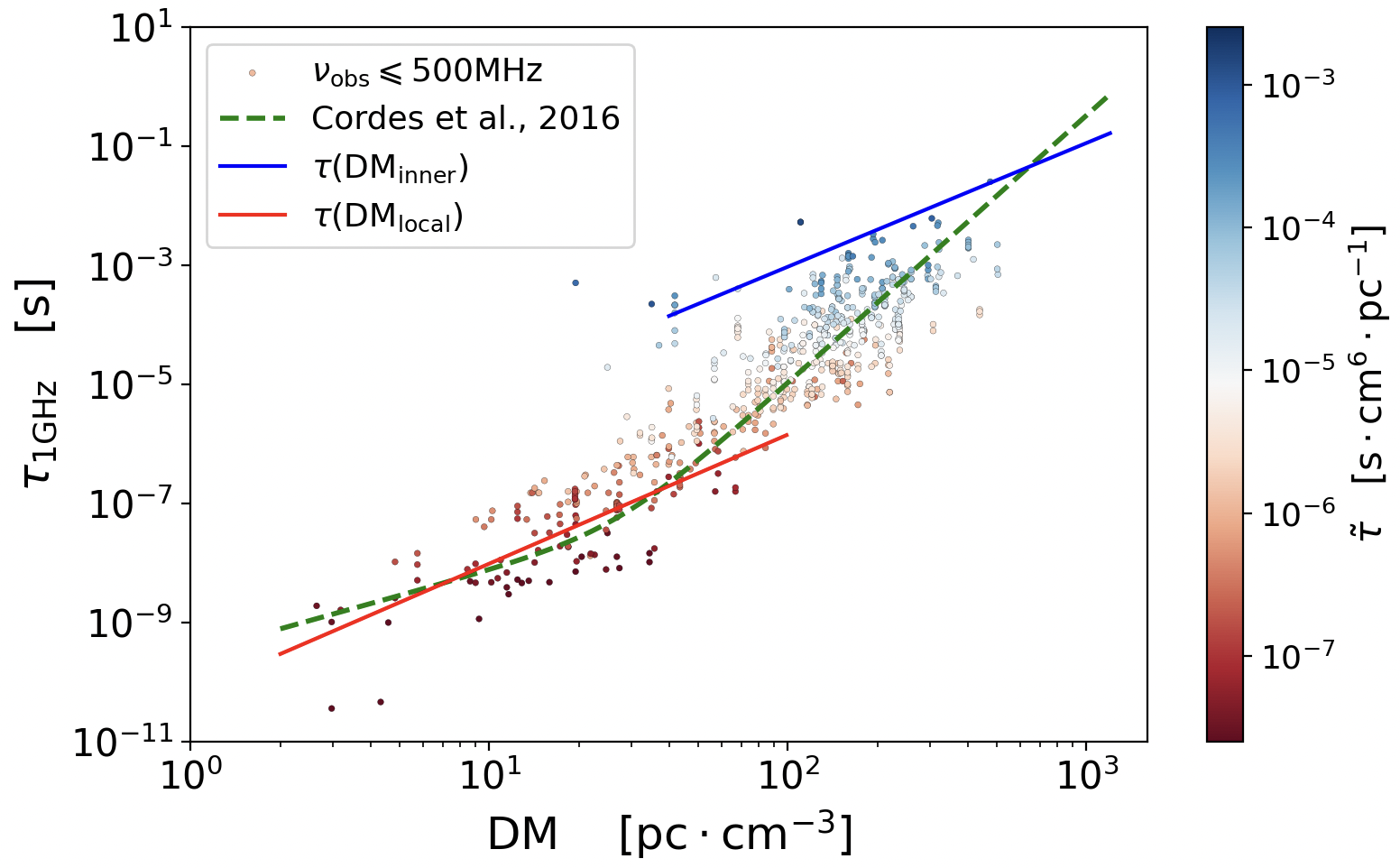}\\
    \includegraphics[width=8.3cm]{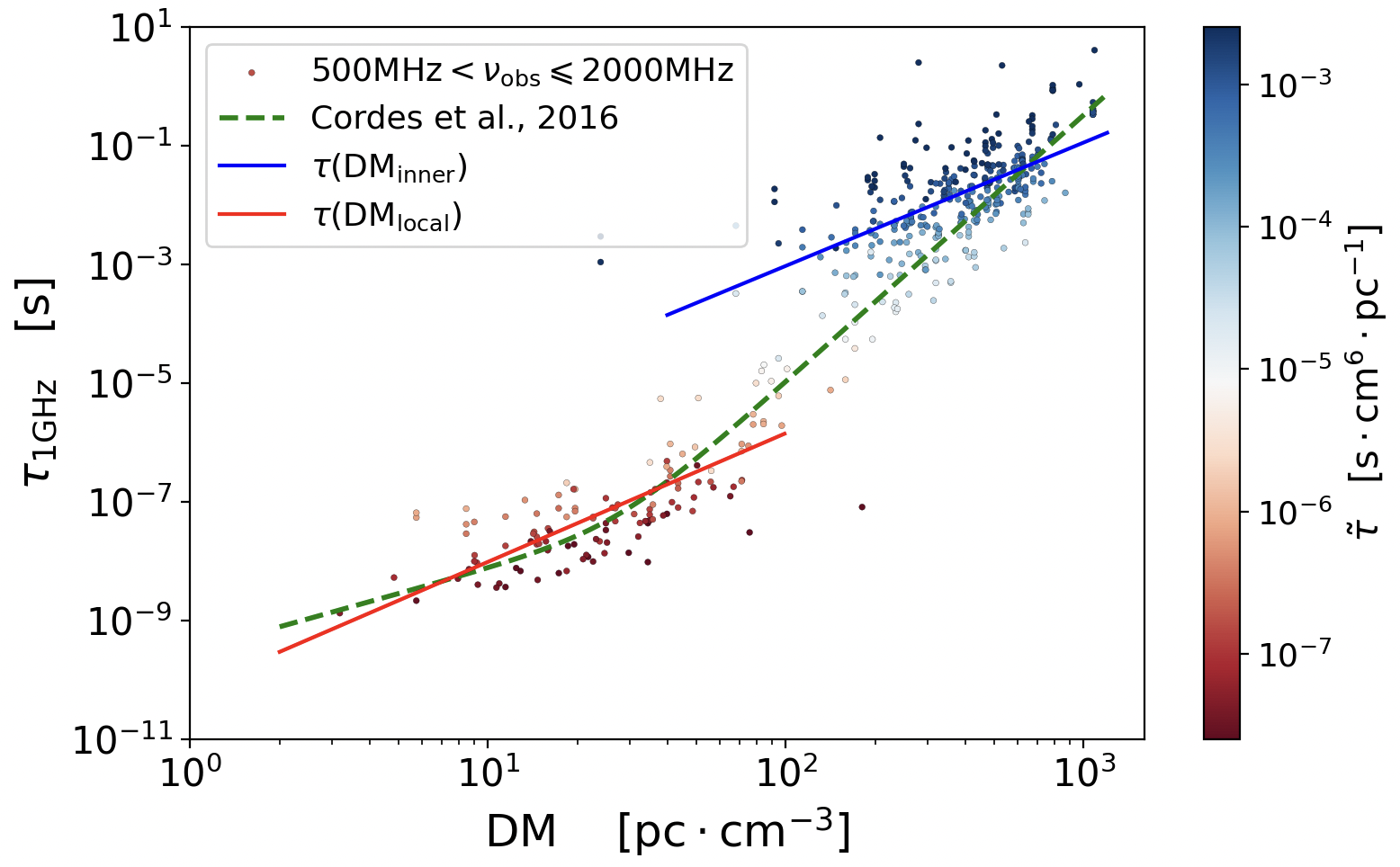}
    \caption{Scatter plots of the relationship between the pulsar scintillation time delay $\tau$ and $\rm DM$ for low-frequency and high-frequency observations. The first panel shows the low-frequency observations below 500 MHz, and the second panel shows the high-frequency observations from 500 MHz to 2000 MHz. It is not difficult to find that there is a piece of missing data in the high-frequency observation data around $\rm DM$ of ${\rm 100 \ pc \ cm^{-3}}$.}
    \label{Data:lhfre}
\end{figure}

\begin{figure}
	\includegraphics[width=8.3cm]{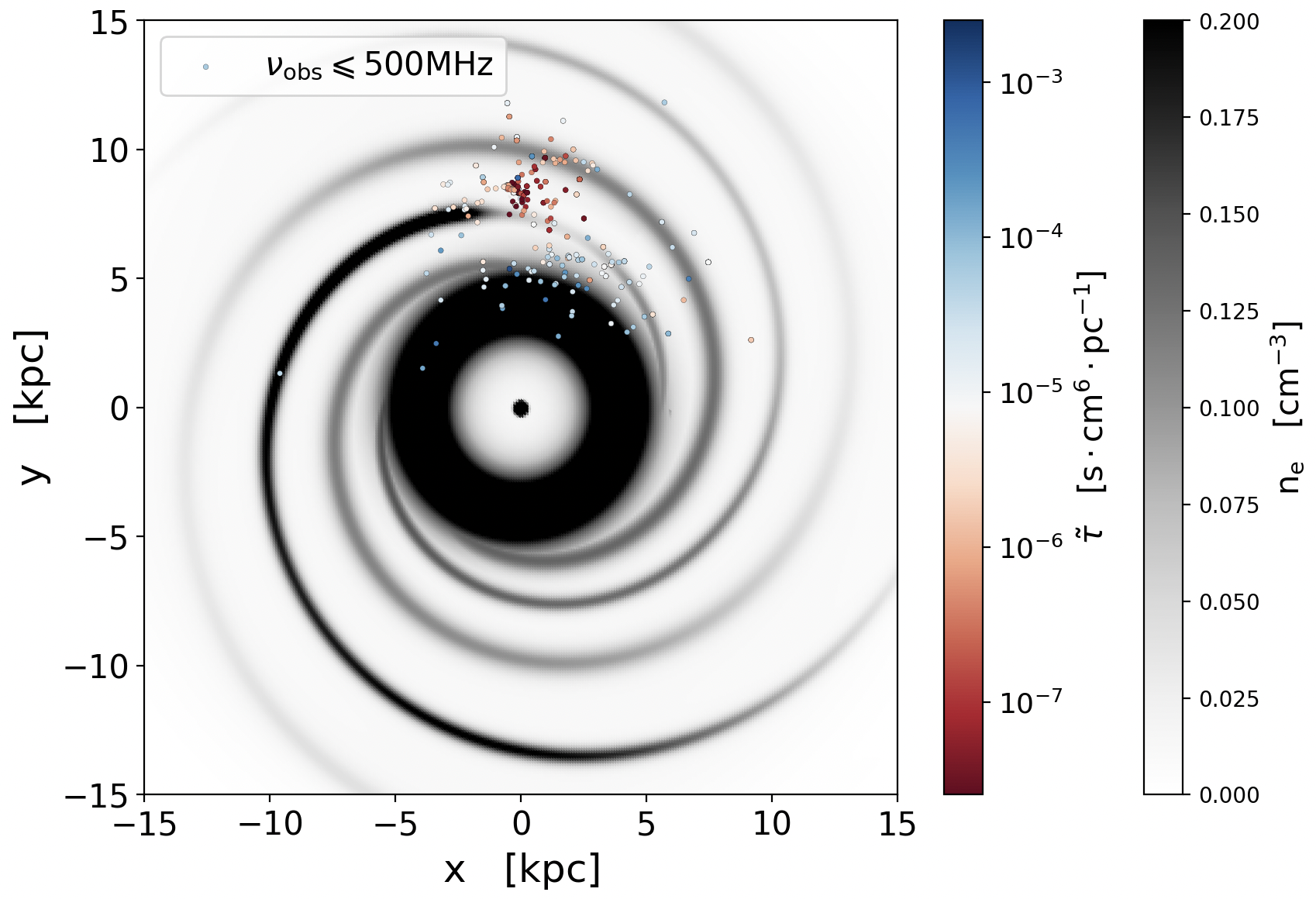}\\
    \includegraphics[width=8.3cm]{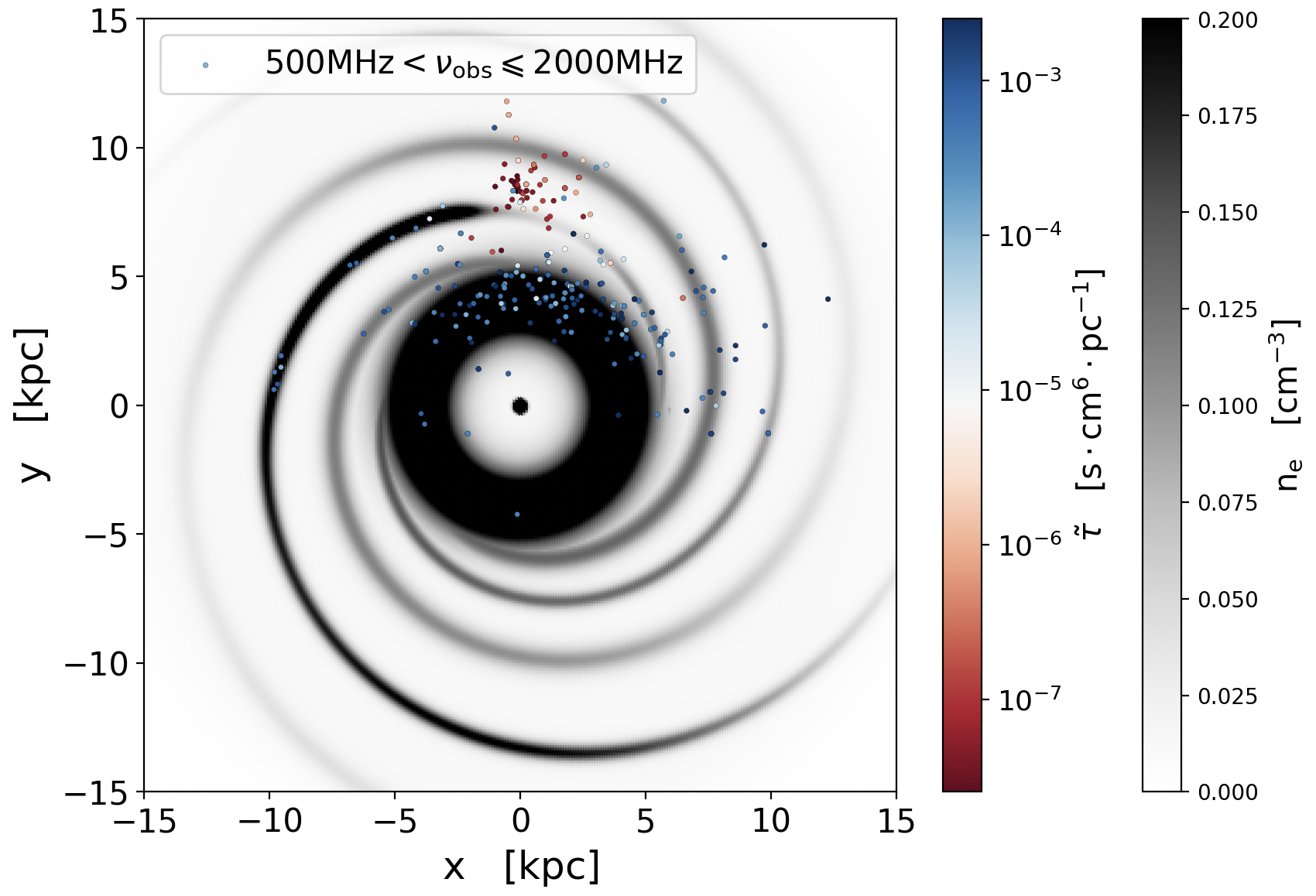}
    \caption{The distribution of low Galactic latitude pulsars in the galaxy after distinguishing between low and high-frequency observations. The first panel shows the low-frequency observations with observation frequencies below 500 MHz, and the second panel shows the high-frequency observations with observation frequencies between 500 MHz and 2000 MHz.}
    \label{sc:lhymw}
\end{figure}

The scattering time $\tau $ is highly dependent on the observing frequency, and their relation is typically taken to have: $\tau _{\rm d}\propto \nu _{\rm obs}^{-\gamma }$, where the scattering index $\gamma$ is related to the power-law spectral index $\alpha$ by the equation $\gamma = -2\alpha /(\alpha -2)$. The value of the index $\alpha$ reflects the different physical properties of the interstellar medium. When the electron density fluctuations in the interstellar medium follow a Kolmogorov turbulence spectrum with $\alpha=11/3$, the scattering index $\gamma=4.4$. On the other hand, a steeper power-law spectrum (also known as Gaussian turbulence spectra) with $\alpha=4$ yields a scattering index $\gamma$ of 4. Several measurements of $\gamma$ values have been conducted so far \citep[e.g.][]{40.2001ApJ...562L.157L, 15.2004A&A...425..569L, 17.2004ApJ...605..759B, 16.2013MNRAS.434...69L, 14.2015MNRAS.449.1570L}, revealing variations in $\gamma$ values among different pulsars and even for the same pulsar at different times. \citet{57.2022A&A...663A.116W} reported independent measurements of $\gamma$ from a Low Frequency Array (LOFAR) data set. Their results show that most of the measured $\gamma$ values are consistent with the predictions of Kolmogorov and Gaussian spectra, except for the PSR J1239+2453 ($\gamma = 3.21 \pm  0.29$) and PSR J2022+2854 ($\gamma= 6.56 \pm 0.24$), for which the $\gamma$ values are too low and too steep, respectively.
\citet{58.2021MNRAS.504.1115O} made observational measurements of a large number of pulsars with the MeerKAT telescope and filtered out pulsars that could lead to inaccurate $\tau$ measurements from the modeling results; they obtained mean and standard deviation of $\left \langle \gamma \right \rangle = 4.0\pm 0.6$. 

\subsection{Low and high-frequency observations}
The pulsar scattering data in our sample are originally taken at a wide range of frequencies from around 40 MHz to beyond 2000 MHz. In this paper, we used $\gamma = 4.0$ to normalize all pulsar scattering measurements to 1 GHz. This extrapolation could, in principle, give rise to large errors in the scattering data of some pulsars. Therefore, we examine the influence of this normalization on our results by splitting the data into sub-samples of low-frequency ($\nu_{\rm obs}\leqslant  {\rm 500MHz}$) and high-frequency (${\rm 500MHz}< \nu_{\rm obs}\leqslant  {\rm 2000MHz}$) observations, respectively. Within each sub-sample, the normalization-led error should be smaller due to a smaller extrapolation range. We then inspect whether the statistics of the sub-samples are consistent with that of the whole sample. Fig.~\ref{Data:lhfre} shows the $\tau-$DM relations for the two sub-samples. Low-frequency (upper panel)  and high-frequency (lower panel)  observations occupy different regions on the $\tau-$DM plane: while the low-frequency data span continuously up to DM$\approx 500$ and $\tau_{\rm 1GHz} \approx 10^{-3}$\,s, the high-frequency data clearly form two populations around $\tau_{\rm 1GHz} \approx 10^{-8}$\,s and $\tau_{\rm 1GHz} \approx 10^{-2}$\,s, respectively. It is thus the high-frequency data that dictated the separation of the pulsar scattering data into local and inner-Galactic populations. 

This is confirmed by Fig.~\ref{sc:lhymw} which shows the Galactic distribution of the scattering data. For the high-frequency data (lower panel), the $\tilde{\tau}$ part (blue) clearly belongs to the inner-Galactic pulsar population, and the low $\tilde{\tau}$ part (red) to the local population. For the low-frequency data (upper panel), however, the boundary between the local and inner-Galactic populations is less clear. 

These apparently different distributions of the low- and high-frequency data arise primarily from an observational selection effect, that different pulsar populations are observed at different frequencies. Not only low-frequency observations of highly-scattered ($\tau_{\rm 1GHz}>10^{-3}$\,s) pulsars are lacking, but also high-frequency observations of the `transitional' pulsars with $\tau_{\rm 1GHz} \sim 10^{-6} - 10^{-4}$\,s. Due to this selection effect, it is still hard to tell whether the underlying distributions of low- and high-frequency data of the same pulsar population are different. Future observations, especially those with ultra-wide-band receivers, will allow us to judge if the normalization of $\tau$ using a fixed $\gamma$ value is well-justified.

\section{Pulsar scattering data}
\label{B} 

\begin{table*}
	\centering
	\caption{Pulsar scattering data and their sources. The full table is available online.}
    \label{tab}
	\begin{tabular}{ccccccccccccc} % four columns, alignment for each
		\hline
		BNAME       & JNAME       & \textit{l}       & \textit{b}       & \textit{d}      & \textit{Z}       & \textit{X}       & \textit{Y}       & DM        & \textit{f}       & tau       & Reference & Parallax \\
        \hline
        B0011+47    & J0014+4746  & 116.497 & -14.631 & 1.776  & -0.449  & 1.538   & 9.267   & 30.405    & 111.0   & 0.004     & 1         & No       \\
        J0023+0923  & J0023+0923  & 111.383 & -52.849 & 1.111  & -0.886  & 0.625   & 8.745   & 14.3      & 1500.0  & 6.2e-09   & 2         & No       \\
        J0026+6320  & J0026+6320  & 120.176 & 0.593   & 6.619  & 0.069   & 5.722   & 11.827  & 245.06    & 325.0   & 0.0476    & 3         & No       \\
        J0026+6320  & J0026+6320  & 120.176 & 0.593   & 6.619  & 0.069   & 5.722   & 11.827  & 245.06    & 610.0   & 0.0075    & 3         & No       \\
        J0030+0451  & J0030+0451  & 113.141 & -57.611 & 0.303  & -0.25   & 0.149   & 8.364   & 4.332772  & 440.0   & 1.236e-09 & 4         & Yes      \\
        J0034-0534  & J0034-0534  & 111.492 & -68.069 & 1.348  & -1.25   & 0.468   & 8.684   & 13.765    & 111.0   & 0.001     & 1         & No       \\
        B0031-07    & J0034-0721  & 110.42  & -69.815 & 1.075  & -1.003  & 0.348   & 8.429   & 10.922    & 44.0    & 0.003     & 1         & Yes      \\
        B0031-07    & J0034-0721  & 110.42  & -69.815 & 1.075  & -1.003  & 0.348   & 8.429   & 10.922    & 1000.0  & 4.229e-09 & 5         & Yes      \\
        B0037+56    & J0040+5716  & 121.452 & -5.567  & 9.804  & -0.945  & 8.324   & 13.391  & 92.5146   & 150.0   & 0.04      & 47        & Yes      \\
        B0037+56    & J0040+5716  & 121.452 & -5.567  & 9.804  & -0.945  & 8.324   & 13.391  & 92.5146   & 200.0   & 0.017     & 54        & Yes      \\
        B0045+33    & J0048+3412  & 122.255 & -28.666 & 4.5    & -2.159  & 3.339   & 10.607  & 39.922    & 111.0   & 0.003     & 1         & No       \\
        B0052+51    & J0055+5117  & 123.621 & -11.576 & 2.857  & -0.567  & 2.331   & 9.85    & 44.013    & 111.0   & 0.007     & 1         & Yes      \\
        \ldots      & \ldots      & \ldots  & \ldots  & \ldots & \ldots  & \ldots  & \ldots  & \ldots    & \ldots  & \ldots    & \ldots    & \ldots   \\
		\hline
	\end{tabular}
\end{table*}

We present here the full set of pulsar scattering data we collected and their references (Table~\ref{tab}). The columns in the table are explained in the following. `BNAME' and `JNAME' represent the name of the pulsar (`JNAME' based on J2000 coordinates), `\textit{l}' and `\textit{b}' are the Galactic longitude and latitude, \textit{d} represent the distance of the pulsar (kpc), `\textit{Z}', `\textit{X}', and `\textit{Y}' represent the specific position of the pulsar in the Galactic coordinate system with the Galactic centre as the origin the Sun located at (0, 8.3, 0; in kpc), `DM' refers to the dispersion measure of the pulsar (pc\,cm$^{3}$), `\textit{f}' refers to the observing frequency of the pulsar (MHz), and "tau" represents the pulse broadening of the pulsar at the observing frequency `\textit{f}' (s). The `Parallax' column indicates whether there are public parallax data \footnote{http://hosting.astro.cornell.edu/research/parallax/}. If `Yes', the `\textit{d}' value is derived from the measured parallax and the `\textit{Z}', `\textit{X}', and `\textit{Y}' values are updated accordingly. Otherwise, the position information of the pulsar comes from the results on the ATNF catalog (version 1.70).  

Reference: (1)\citet{12.2007A&AT...26..597K}; (2)\citet{20.2016ApJ...818..166L}; (3)\citet{5.2019ApJ...870....8S}; (4)\citet{21.2001A&A...368.1055N}; (5)\citet{22.1986ApJ...311..183C}; (6)\citet{23.1998MNRAS.297..108J}; (7)\citet{14.2015MNRAS.449.1570L}; (8)\citet{24.1999ApJS..121..483B}; (9)\citet{25.2015ApJ...804...23K}; (10)\citet{26.1986AuJPh..39..433A}; (11)\citet{27.1985ApJ...288..221C}; (12)\citet{28.2001A&A...370..586M};(13)\citet{29.1997MNRAS.290..260R}; (14)\citet{30.1977ARA&A..15..479R}; (15)\citet{31.1970ApL.....6..147A}; (16)\citet{32.1982MNRAS.201.1119R}; (17)\citet{13.1992Natur.360..137P}; (18)\citet{33.1992MNRAS.255..401J}; (19)\citet{34.2000ApJ...531..345G}; (20)\citet{35.1991MNRAS.252...13C}; (21)\citet{16.2013MNRAS.434...69L}; (22)\citet{36.2015MNRAS.454.2517L}; (23)\citet{37.1990BAAS...22.1244W}; (24)\citet{38.1992MNRAS.254..177C}; (25)\citet{39.2017MNRAS.468.2526B}; (26)\citet{40.2001ApJ...562L.157L}; (27)\citet{41.1985MNRAS.212..975M}; (28)\citet{42.2007A&A...462..703C}; (29)\citet{43.2000MNRAS.312..698L}; (30)\citet{44.2012ApJ...746...63C}; (31)\citet{45.1981AJ.....86..418R}; (32)\citet{17.2004ApJ...605..759B}; (33)\citet{46.2013ApJ...772...50N}; (34)\citet{47.1988AIPC..174..217D}; (35)\citet{48.1990ApJ...349..245C}; (36)\citet{49.1995ApJ...445..756C}; (37)\citet{50.1988Natur.331...53F}; (38)\citet{51.2004ApJ...612..389H}; (39)\citet{52.2011ApJ...739...39A}; (40)\citet{10.2022arXiv220708756S};(41)\citet{59.2021MNRAS.505.4468G}; (42)\citet{58.2021MNRAS.504.1115O};(43)\citet{60.2021ApJ...917...67C}; (44)\citet{61.2022ApJ...924..135P};(45)\citet{62.2021ApJ...915...15J}; (46)\citet{63.2022A&A...667A..79D};(47)\citet{69.2017MNRAS.470.2659G}; (48)\citet{70.2017ApJ...846..104K};(49)\citet{71.1998MNRAS.301..235G}; (50)\citet{72.1998MNRAS.297...28D};(51)\citet{73.1994PhDT.......272L}; (52)\citet{74.1998MNRAS.295..280M};(53)\citet{75.2015ApJ...808..156S}; (54)\citet{76.2019ApJ...878..130K}.

\section{Posterior distributions of fitting parameters}
\label{C}

\begin{figure}
	\includegraphics[width=8.3cm]{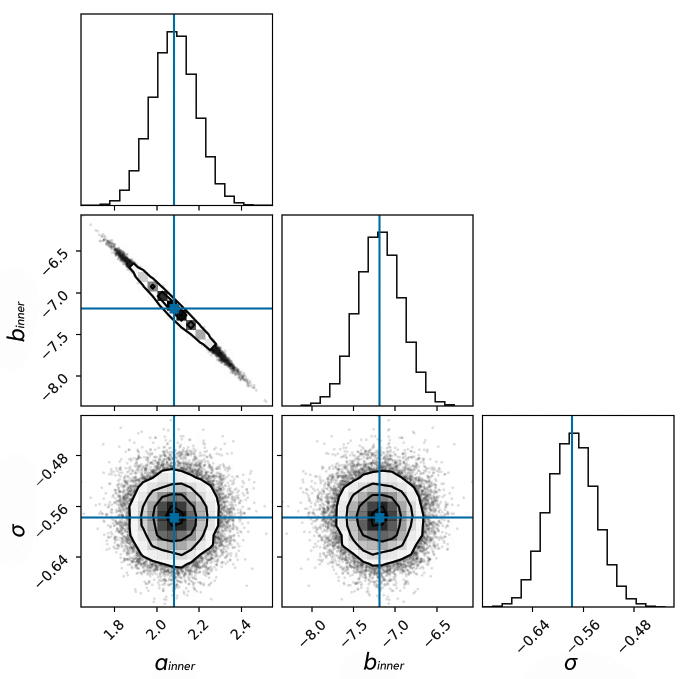}\\
    \includegraphics[width=8.3cm]{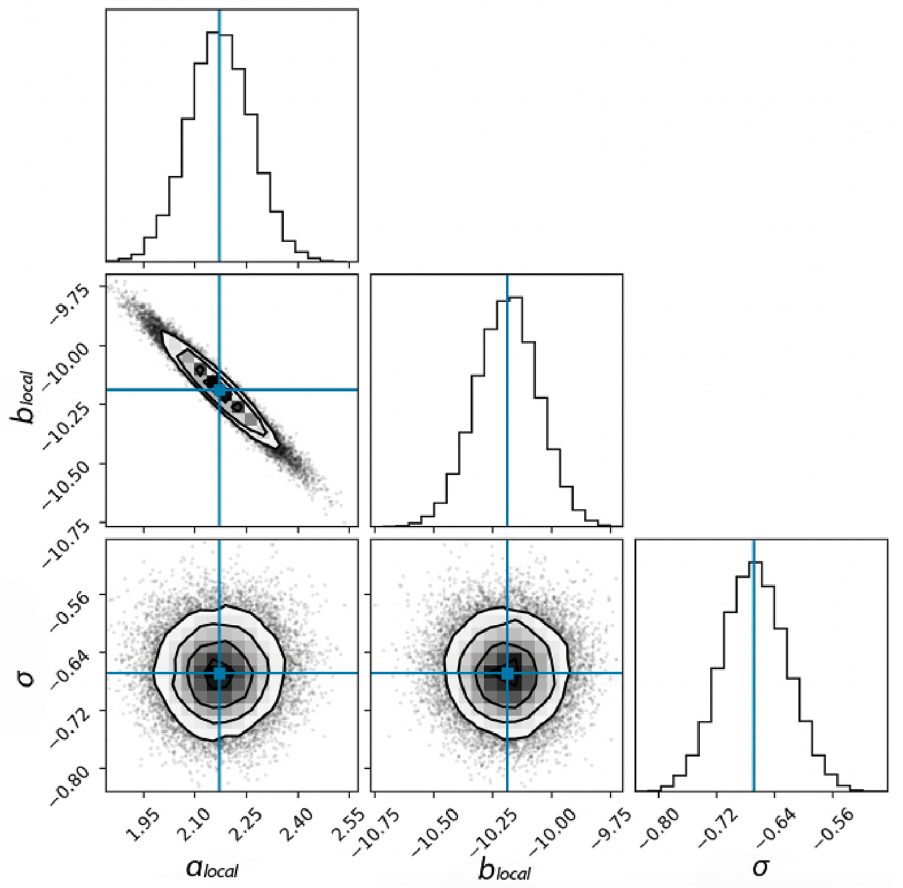}
    \caption{Posterior probability distributions for scattering time-scale $\tau$ at 1 GHz against DM for two pulsar populations, where $\tau (\rm DM)={\rm DM}^{\it a}\times 10^{\it b}$, and these two graphs show the `inner' pulsar population (top) and the `local' pulsar population (bottom). The $\tau$-DM relation with the best-fit parameters is shown in Fig.~\ref{Data:ourresult}.}
    \label{resmcmc}
\end{figure}

\begin{figure}
	\includegraphics[width=8.3cm]{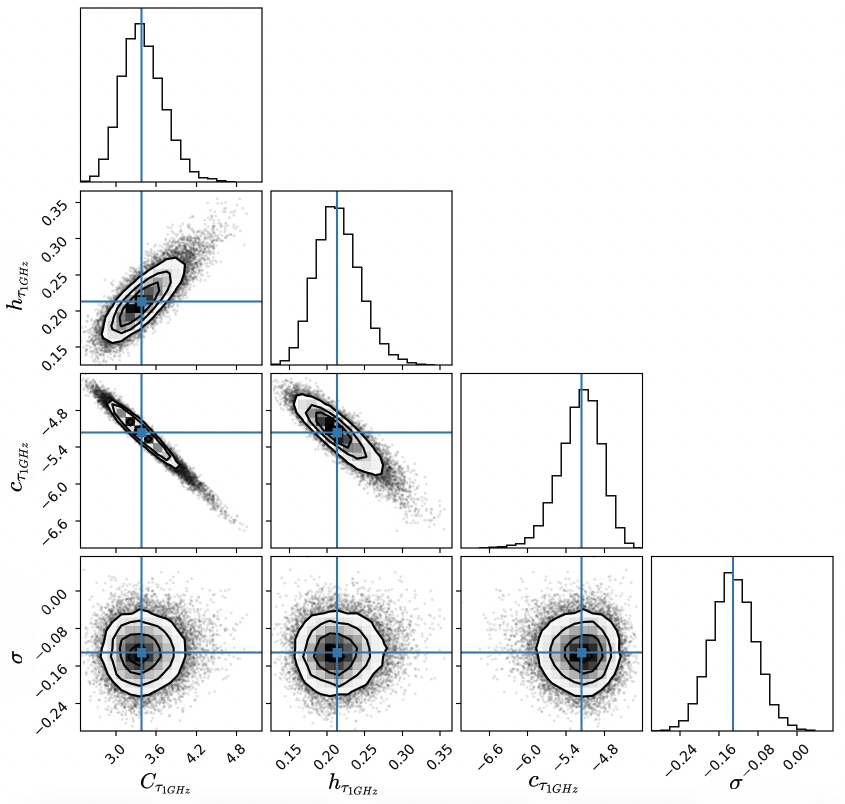}\\
    \includegraphics[width=8.3cm]{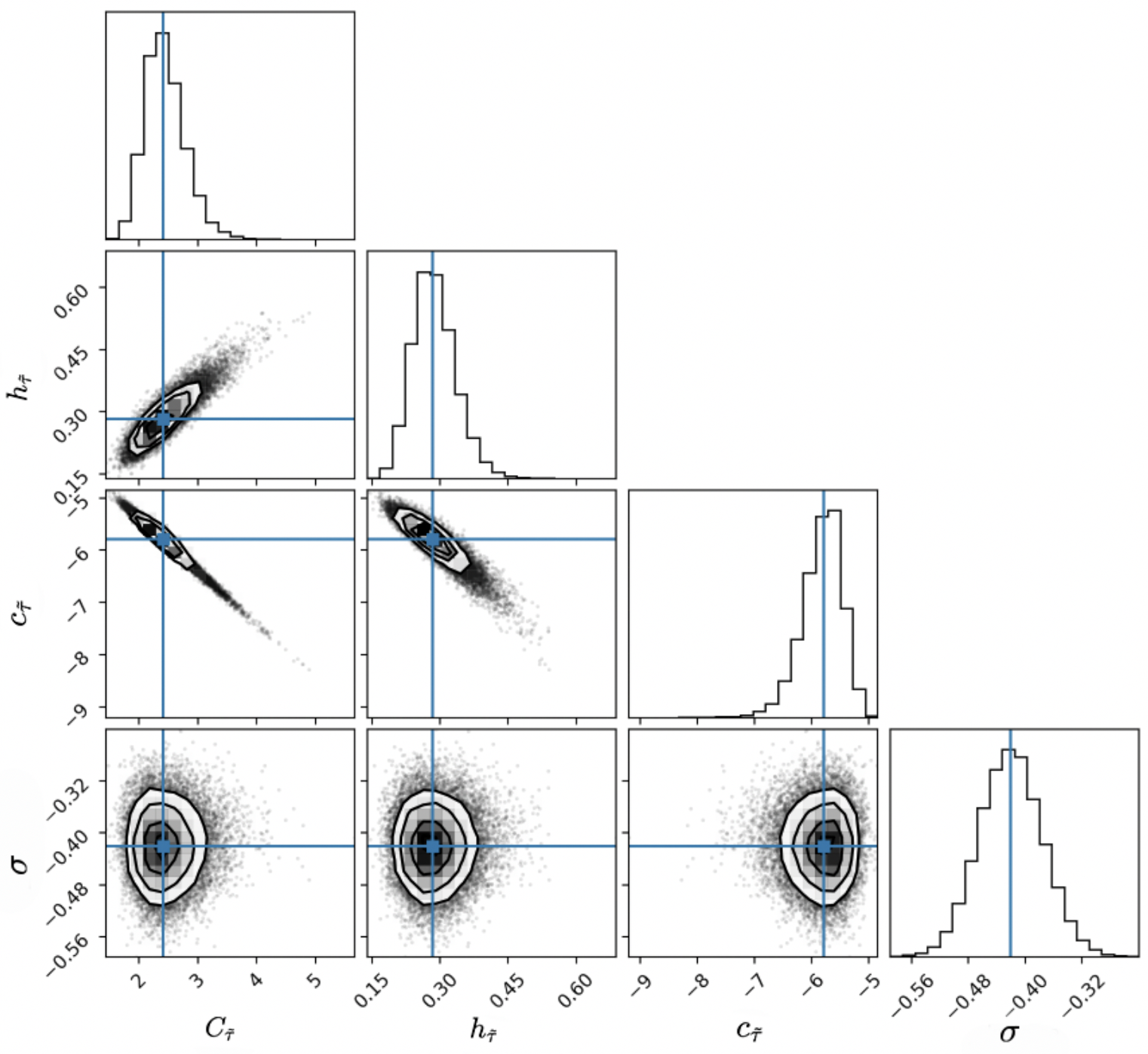}
    \caption{Posterior probability distributions for scattering time-scale $\tau$ at 1 GHz (top) and reduced scattering strength $\tilde \tau$ (bottom) against $\left |\rm{Z} \right |$, where $\mathrm{log \mathit{K} = \mathit{C} sech(\frac{\left |\rm{Z} \right |}{\mathit{h}})+\mathit{c}}$, $K$ is $\tau_{\rm 1GHz}$ or $\tilde \tau$. The results with the best-fit parameters are shown in Fig.~\ref{hifit}.}
    \label{himcmc}
\end{figure}

Figures \ref{resmcmc} and \ref{himcmc} show the posterior distributions between different parameters. In these figures, blue lines indicate mean values and contours show 1$\sigma$, 2$\sigma$ and 3$\sigma$ confidence levels. Fig.~\ref{resmcmc} shows the distributions resulting from the $\tau$-DM relation for the two pulsar populations discussed in Section~\ref{sc:relation} and Fig.~\ref{himcmc} shows the distributions resulting from the scattering scale height of the inner disc discussed in Section~\ref{sc:hi}. 

% Don't change these lines
\bsp	% typesetting comment
\label{lastpage}
\end{document}